\documentclass[a4papaer,twocolumn,amsmath,amssymb,prl,10pt,nofootinbib,superscriptaddress]{revtex4}
\usepackage{ graphicx, float,amsmath, amsmath, amssymb}
\usepackage[export]{adjustbox}
\newcommand{\qsubrm}[2]{{#1}_{\scriptscriptstyle{\textrm{#2}}}}
\newcommand{\qsuprm}[2]{{#1}^{\scriptscriptstyle\textrm{#2}}}
\def\be{\begin{equation}}
\def\ee{\end{equation}}
\def\bea{\begin{eqnarray}}
\def\eea{\end{eqnarray}}
\def\bse{\begin{subequations}}
\def\ese{\end{subequations}}

\usepackage[breaklinks, colorlinks, citecolor=blue]{hyperref}
\begin{document}
\title{Some Clarifications on the Duration of Inflation in Loop Quantum Cosmology}

\author{Boris Bolliet}
\email[Corresponding author: ]{boris.bolliet@lpsc.in2p3.fr}
\affiliation{%
Laboratoire de Physique Subatomique et de Cosmologie, Universit\'e Grenoble-Alpes, CNRS/IN2P3\\
53, avenue des Martyrs, 38026 Grenoble cedex, France
}
\author{Aur\'elien Barrau}%
\affiliation{%
Laboratoire de Physique Subatomique et de Cosmologie, Universit\'e Grenoble-Alpes, CNRS/IN2P3\\
53, avenue des Martyrs, 38026 Grenoble cedex, France
}
\author{Killian Martineau}%
\affiliation{%
Laboratoire de Physique Subatomique et de Cosmologie, Universit\'e Grenoble-Alpes, CNRS/IN2P3\\
53, avenue des Martyrs, 38026 Grenoble cedex, France
}
\author{Flora Moulin}
\affiliation{%
Laboratoire de Physique Subatomique et de Cosmologie, Universit\'e Grenoble-Alpes, CNRS/IN2P3\\
53, avenue des Martyrs, 38026 Grenoble cedex, France
}
\date{\today}
\begin{abstract} 
The prediction of a phase of inflation whose number of e-folds is constrained is an important feature of loop quantum cosmology. This work aims at giving some elementary clarifications on the role of the different hypotheses leading to this conclusion. We show that the duration of inflation does not depend significantly on the modified background dynamics in the quantum regime.
 \end{abstract}
\maketitle

Loop quantum gravity (LQG) is a  nonperturbative and background-independent quantization of general relativity (GR). It relies on the Sen-Ashtekar-Barbero variables, that is SU(2) valued connections and conjugate densitized triads. The quantization is obtained using holonomies of the connections and fluxes of the densitized triads. Loop quantum cosmology (LQC) is an effective theory based on a symmetry reduced version of LQG. In LQC, the big bang is believed to be replaced by a  bounce due to repulsive quantum geometrical effects (see \cite{2011CQGra..28u3001A} for a review). For the flat homogeneous and isotropic background cosmology that we consider in this work, the effective LQC-modified Friedmann equation is
\begin{equation}
 H^2=\frac{\rho}{3}\left(1-\frac{\rho}{\qsubrm{\rho}{B}}\right),  \label{fried}
 \end{equation}
where $H\equiv(\dot{a}/a)$ is the Hubble parameter, $\rho$ is the total energy density and $\qsubrm{\rho}{B}$ is the critical density at the bounce (expected to be of the order of the Planck density). The dot refers to a coordinate time derivative.



We assume that the dominating energy component in the early universe is a scalar field $\phi$, with potential  $V=\tfrac{1}{2}m^2\phi^2$. The total energy density  can be written as $\rho = \tfrac{1}{2}\dot{\phi}^2 + V$. 
Based on cosmic microwave background (CMB) measurements and under most reasonable assumptions for the length of \textit{observable inflation} (between horizon exit of the pivot scale and the end of the inflationary phase), one obtains $m\simeq10^{-6}\qsubrm{m}{Pl}$. The equation of motion for the scalar field is
\be
\ddot{\phi}+3H\dot{\phi}+m^2\phi=0.\label{eq:KGE}
\ee

There are different ways to statistically estimate the duration of inflation in this framework.\\

At a fixed energy density, $\qsubrm{\rho}{0}$, one can first ask the following question: for a given number of e-folds $N$, what is the fraction of trajectories, \textit{i.e.} solutions to Eq.\eqref{eq:KGE}, that lead to a phase of slow-roll inflation lasting more than $N$ e-folds?
It should be noticed that the set of trajectories can be parametrized by $\{\qsubrm{a}{0},\qsubrm{\phi}{0}\}$. As the energy density has been fixed, the initial time derivative of the scalar field, $\qsubrm{\dot{\phi}}{0}$, is determined in terms of $\qsubrm{\rho}{0}$ and  $\qsubrm{\phi}{0}$. This also implies that $\qsubrm{\phi}{0}$ can only take values within a finite interval, ranging from $-(\sqrt{2\qsubrm{\rho}{0}}/m)$ to $(\sqrt{2\qsubrm{\rho}{0}}/m)$. In a flat universe, the value of the scale factor has no physical meaning. 
The number of e-folds of inflation depends on $\qsubrm{\phi}{0}$ but not on $\qsubrm{a}{0}$: $N=N(\qsubrm{\phi}{0};m,\qsubrm{\rho}{0})$. So the fraction of trajectories that achieve a phase of inflation lasting more than $N$ e-folds can be written as $\mu=(m\Delta\qsubrm{\phi}{0})/(2\sqrt{2\qsubrm{\rho}{0}})$, where $\Delta\qsubrm{\phi}{0}$ is the range of initial values of the scalar field that yields the required inflationary phase. It is then necessary to evaluate $\mu$ as a function of $N$. There are two cases in which this can be done analytically: (i) at \textit{low energy}, $\qsubrm{\rho}{0}\ll m^2$, and (ii) at \textit{high energy} $\qsubrm{\rho}{0}\gg m^2$.  At low energy, the calculation of Gibbons and Turok of the probability for inflation can be used to show that \cite{2008PhRvD..77f3516G}
\be
\mu(N) = \mathcal{C}mN^{-\frac{1}{2}}\exp(-3N)\{1+1/(6N)\},
\ee
where $\mathcal{C}$ is a numerical factor that does not depend on $m$ or $\qsubrm{\rho}{0}$. For $N\simeq60$ e-folds,  as required to explain the CMB temperature anisotropy, this leads to $\mu(N)\ll1$. At high energy, one reaches the opposite conclusion. In this case, one can compute $\Delta\qsubrm{\phi}{0}$ as follows. For a massive quadratic potential the total number of e-folds of inflation can be expressed in terms of the amplitude of the scalar field at the start of the inflationary phase, $\qsubrm{\phi}{I}$, as $N\approx (\qsubrm{\phi}{I}^2/4)$. In turn, $\qsubrm{\phi}{I}$ can be expressed in terms of the initial value of the scalar field as
\be
\qsubrm{\phi}{I}=\qsubrm{\phi}{0}+\mathrm{sgn}(\qsubrm{\dot{\phi}}{0})\sqrt{(2/3)}\mathrm{Arcsinh}\left({\Gamma\sqrt{2/\ln(z)}}\right), \label{eq:maxphi}
\ee
with $z\equiv 8\Gamma^2\exp(\sqrt{6}\qsubrm{\phi}{0})$ and $\Gamma\equiv\sqrt{3\qsubrm{\rho}{0}}/m$. This formula for the amplitude of the scalar field at the start of inflation is valid in LQC, with the modified Friedmann equation given by Eq. \eqref{fried}. For the standard flat FLRW dynamics, without LQC modifications, the analytical calculations suggest that at the start of inflation the scalar field reaches a maximum value given by \eqref{eq:maxphi} minus $ (\ln 2/\Gamma)$. In both cases, we find that the range of values of $\qsubrm{\phi}{0}$ that do not yield an inflationary phase longer than $N$ e-folds is an interval of size $4\sqrt{N}$ centered on $\qsubrm{\phi}{0}=0$. Hence,
\be
\mu(N)=1-m\sqrt{(2N/\qsubrm{\rho}{0})},
\ee
and $\mu(60)\simeq0.99999$, which means that all but a tiny fraction of the possible trajectories do not go through a long inflationary phase.
It might be tempting to interpret $\mu$ as a probability measure. This is however not that simple. The phase space of the flat FLRW universe presents a serious ambiguity: the Liouville measure  is proportional to the scale factor and the scale factor can be rescaled arbitrarily. In addition, as explained just before, $\mu$ depends on the choice of the surface of initial data.  More importantly, the fundamental question to ask is: is there a variable on which a flat (or at least known) probability distribution function (PDF) can be assigned? Assuming implicitly that the initial values of the field should have a flat PDF is arbitrary.\\

\begin{figure*}
\includegraphics[width=12cm]{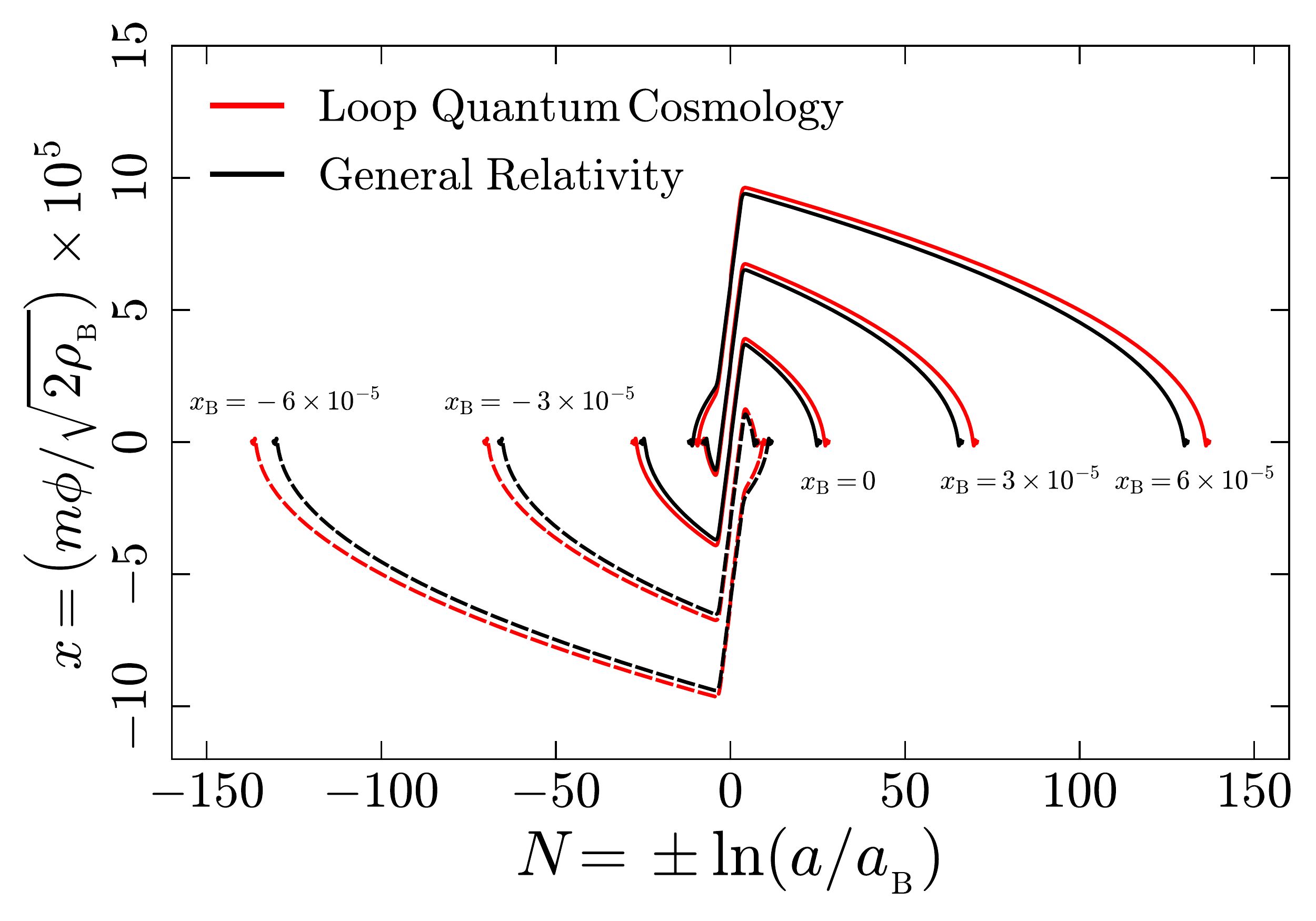}
\caption{Evolution of the potential energy parameter in the GR-like scenario (black) compared to loop quantum cosmology (red), for different values of $\qsubrm{x}{B}$, linearly distributed between $-10^{-6}$ and $10^{-6}$.  }
\label{xlambda}
\end{figure*}
In  \cite{AS2011} it was argued  that the two first issues mentioned above can be solved in LQC. It was indeed claimed that the scale factor can be rigorously factored out of the Liouville measure, and that the bounce provides a preferred choice for the surface of initial data. In this study, following \cite{Linsefors:2013cd,Linsefors:2014tna} we choose a different perspective. We decide, the other way round, to set initial conditions in the remote past of the contracting branch, when the Universe is classical and well understood ($\qsubrm{\rho}{0}\ll\qsubrm{\rho}{B}$). This is not only technically justified but also conceptually necessary if the bounce has to be taken seriously in a causal way. Still, we naturally choose a time which is close enough to the bounce so that it is reasonable to assume a scalar field as the main component of the Universe. The phase of the oscillations of the scalar field in the contracting branch  is an obvious variable to which a flat PDF can be assigned \cite{Linsefors:2013cd}. In addition, the key point is that this PDF is preserved over time (as long as one remains in the classical phase when the field oscillates).
The numerical analysis of \cite{Linsefors:2013cd} shows that at fixed $\qsubrm{\rho}{0}$, nearly all possible initial values for the scalar field, $\qsubrm{\phi}{0}$, yield an inflationary phase whose number of e-folds is peaked around $N=142$ e-folds (with $\qsubrm{\rho}{B}=0.41\qsubrm{m}{Pl}^4$). 

\begin{figure*}
\includegraphics[height=6.cm]{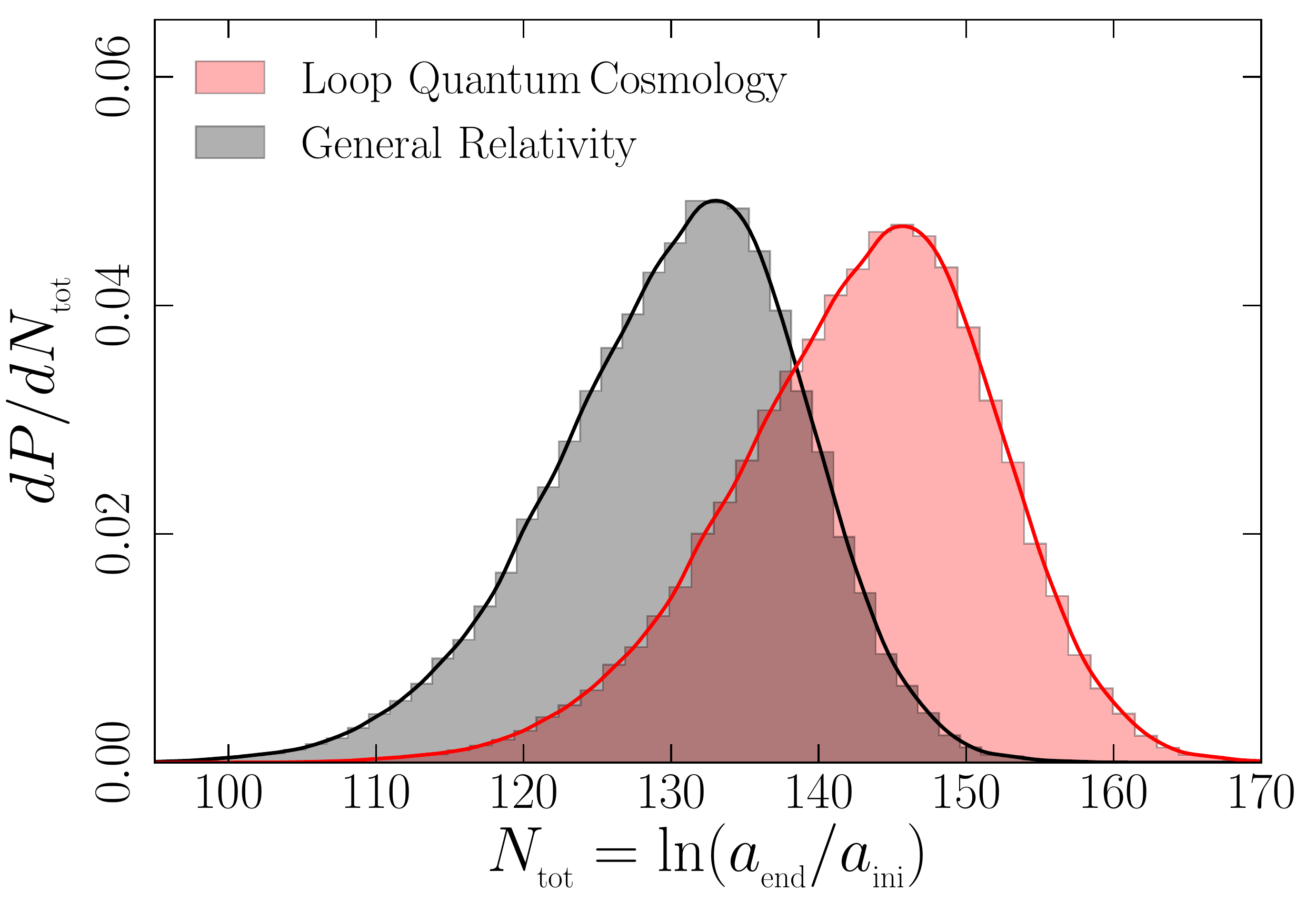}\includegraphics[height=6.cm]{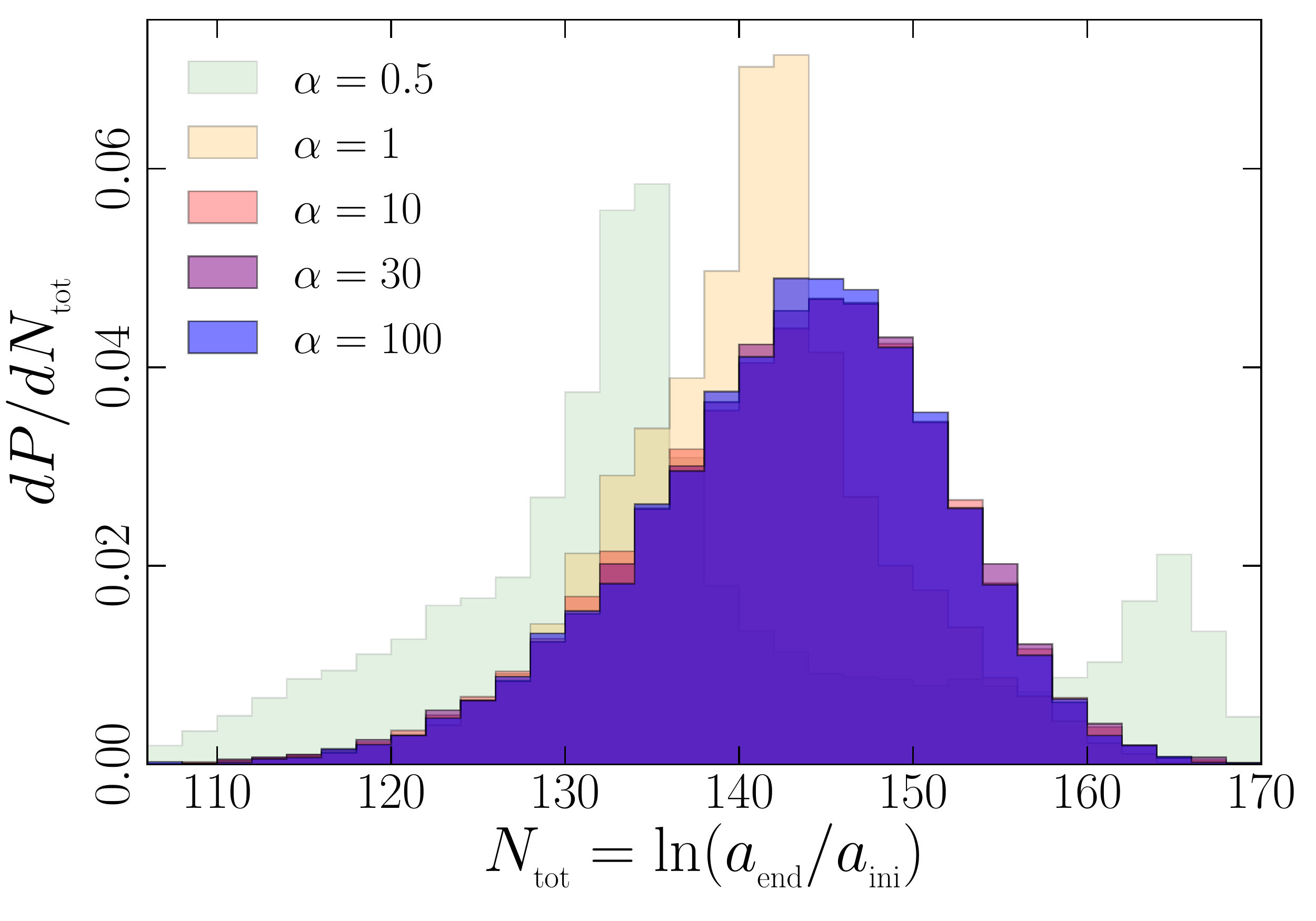}
\caption{Probability distribution of the number of inflationary e-folds. On the left panel, the black histograms corresponds to a `GR' like dynamics (using the standard Friedmann equation throughout the evolution). The red histogram is the prediction of loop quantum cosmology. With the standard Friedmann equation the most likely value is $\qsubrm{N}{tot}=133$, while in LQC we find $\qsubrm{N}{tot}=145$. The right panel shows that the probability density function does not depend on the value of the energy density as long as the surface of initial data is set at $\rho\ll\qsubrm{\rho}{Pl}$. The different histograms are labeled by $\alpha=\sqrt{\qsubrm{\rho}{Pl}/\rho}$. The probability density function converges as soon as $\alpha$ becomes larger than 10.} 
\label{pdf-lqc}
\end{figure*}

\begin{figure}
\includegraphics[height=5.5cm]{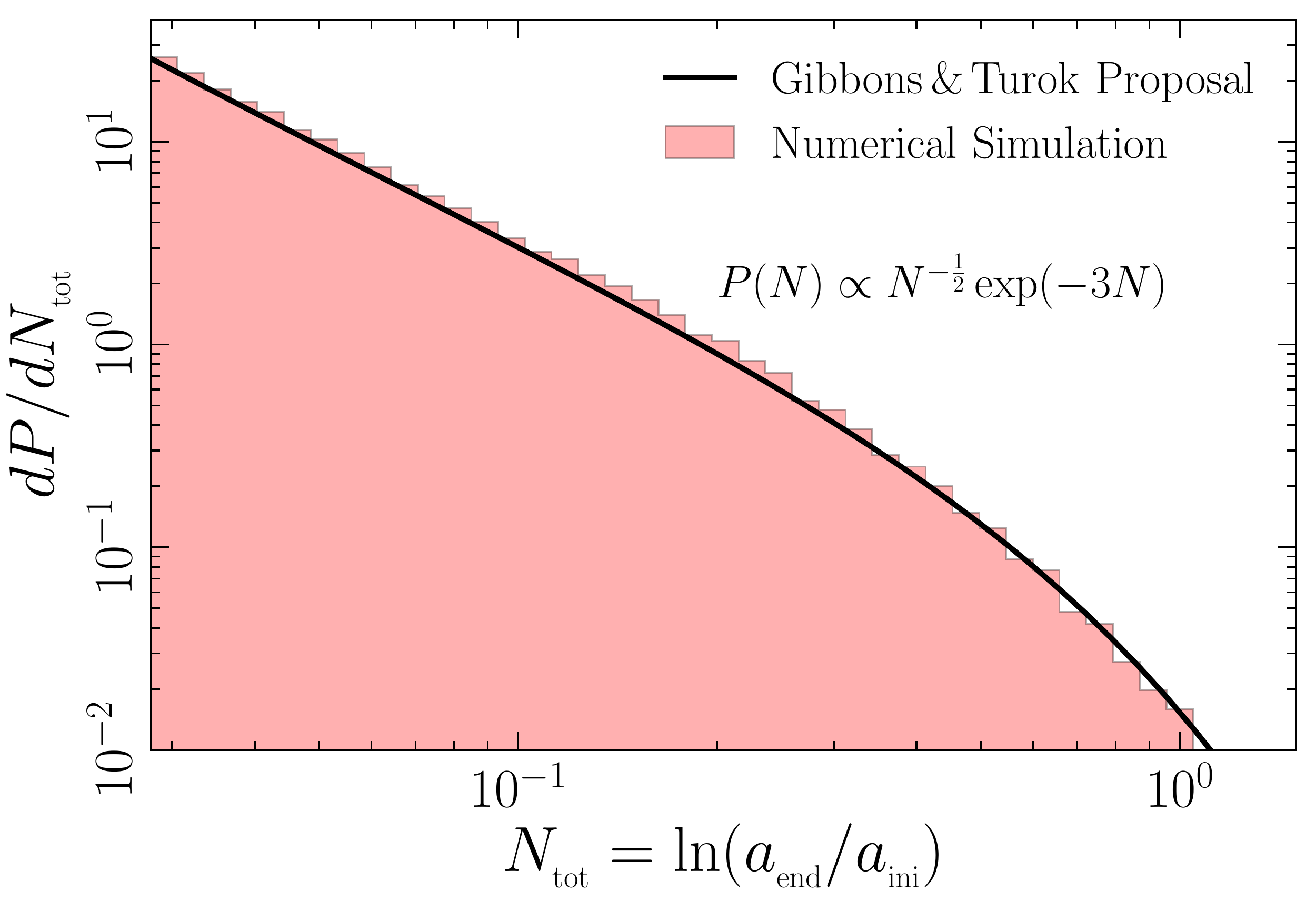}
\caption{Probability distribution of the number of  e-folds, when the surface of initial data is set at low energy density, in a agreement with the result of Gibbons\&Turok.} 
\label{pdf}
\end{figure}

The procedure to derive this result is simple: 
\begin{itemize}
\item Consider an initial energy density $\qsubrm{\rho}{0}=\qsubrm{\rho}{Pl}/\alpha^2$, with $\alpha$  large large enough so that the evolution starts in the remote past of the contracting phase. 
\item Choose an initial value for the scalar field and its time derivative by a random sampling of the phase $\qsubrm{\theta}{0}$ between $0$ and $2\pi$, where  $\qsubrm{\theta}{0}$ is defined such as $\qsubrm{\phi}{0}=\sqrt{\frac{2}{3}}\frac{\Gamma}{\alpha}\sin\qsubrm{\theta}{0}$.
\item Solve the dynamics, across the bounce, until the end of slow-roll inflation in the expanding branch.
\item For each  $\qsubrm{\theta}{0}$, collect the corresponding number of e-folds. 
\end{itemize}
Finally, one can produce the associated histogram which, in a probabilistic interpretation, is the PDF for the number of e-folds. This is illustrated on the right panel of Fig. \ref{pdf} where we also present the PDFs for several initial energy densities corresponding to different values of $\alpha\equiv\sqrt{\qsubrm{\rho}{Pl}/\qsubrm{\rho}{0}}$ in order to show that for large values of $\alpha$ the PDF becomes independent of the initial energy density, as explained analytically in \cite{Linsefors:2013cd}. Interestingly, the peakedness of the PDF can be understood as follows. The calculation Gibbons and Turok is often considered controversial in standard cosmology because they somehow set ``initial conditions" for the {\it final} state. However, in the case of a bouncing Universe it implies that almost none of all the possible trajectories, starting at low energy in the {\it contracting} branch, have a significant phase of pre-bounce exponential contraction, that is of so-called \textit{deflation}. 
A trajectory with deflation in the contracting phase leading to $(\qsubrm{\phi}{B}, \qsubrm{\dot{\phi}}{B})$ can be identified with a trajectory with inflation in the expanding phase with $(\qsubrm{\phi}{B}, -\qsubrm{\dot{\phi}}{B})$. Equation \eqref{eq:maxphi} can be used to calculate the value of the scalar field at the bounce corresponding to the trajectory with no deflation. One simply has to solve Eq. \eqref{eq:maxphi} with respect to $\qsubrm{\phi}{B}$ for $\qsubrm{\phi}{I}=0$ and $\qsubrm{\dot{\phi}}{B}<0$. In the limit of large $\Gamma$, the solution is well approximated by
\be
\qsubrm{\phi}{B}^{\scriptscriptstyle{\mathrm{GT}}}\equiv\sqrt{(2/3)}\ln\left(2\Gamma/\sqrt{\ln\Gamma}\right).\label{eq:gt}
\ee
This can then be inserted back into Eq. \eqref{eq:maxphi}, with $\qsubrm{\dot{\phi}}{B}>0$, in order to obtain the value of the field at the start of inflation in the expanding phase and the corresponding number of e-folds of inflation. With the standard values for $m$ and $\qsubrm{\rho}{B}$, this calculation yields $N=142$, in excellent agreement with the numerics (Fig. \ref{pdf-lqc}). Moreover, a closer look at Gibbons and Turok's PDF for the number of e-folds suggests that most trajectories starting in the remote past have less than one e-folds of deflation, see Fig. \ref{pdf}. This means that nearly all trajectories end up with a value of $\qsubrm{\phi}{B}$ that belongs to an interval of size $\Delta\qsubrm{\phi}{B}\approx4$ centered around $\qsubrm{\phi}{B}^{\scriptscriptstyle{\mathrm{GT}}}$. In terms of number of e-folds this translates into $\Delta N\approx 4\sqrt{N}$, also in agreement with the numerics as can be seen on Fig. \ref{pdf-lqc}.\\ 

\begin{figure*}
\includegraphics[width=15.cm]{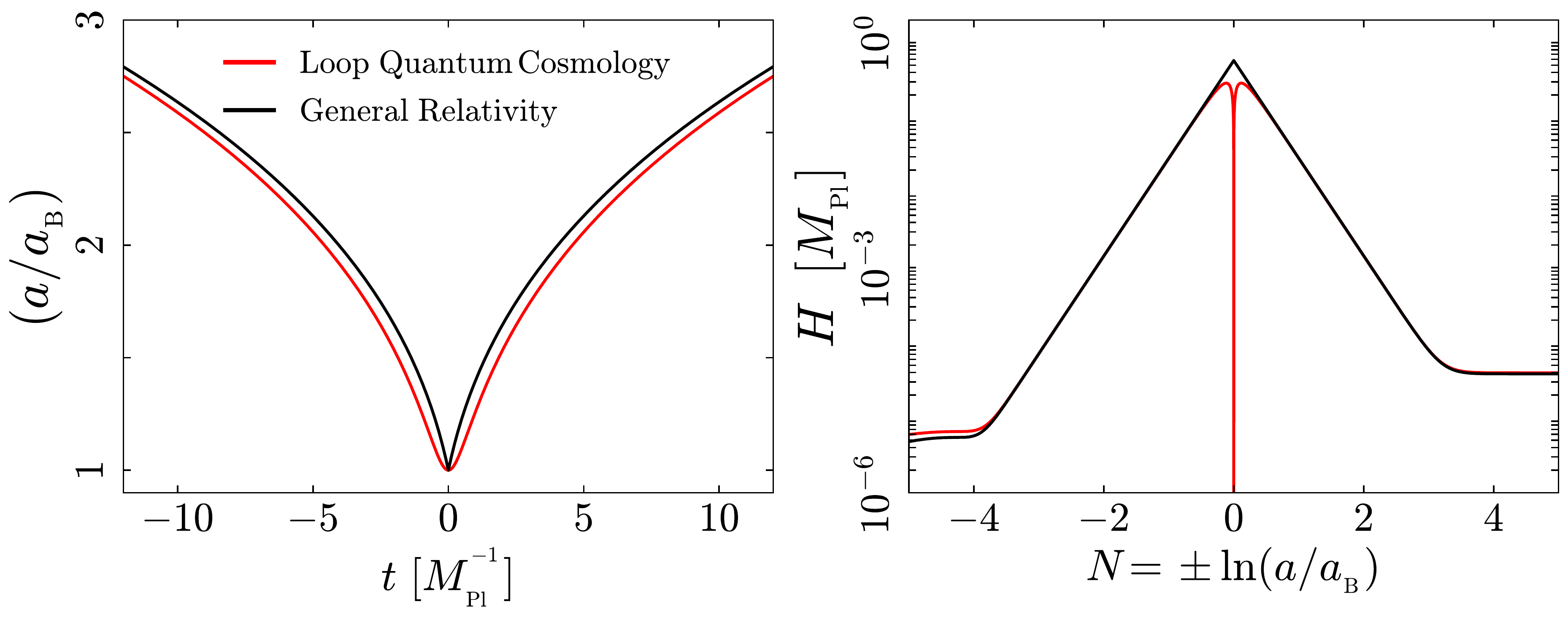}
\caption{Evolution of the scale factor (left) and the Hubble parameter (right) in loop quantum cosmology (red) and in the GR-like scenario (black).} 
\label{aandH}
\end{figure*}

\begin{figure}
\includegraphics[width=8cm]{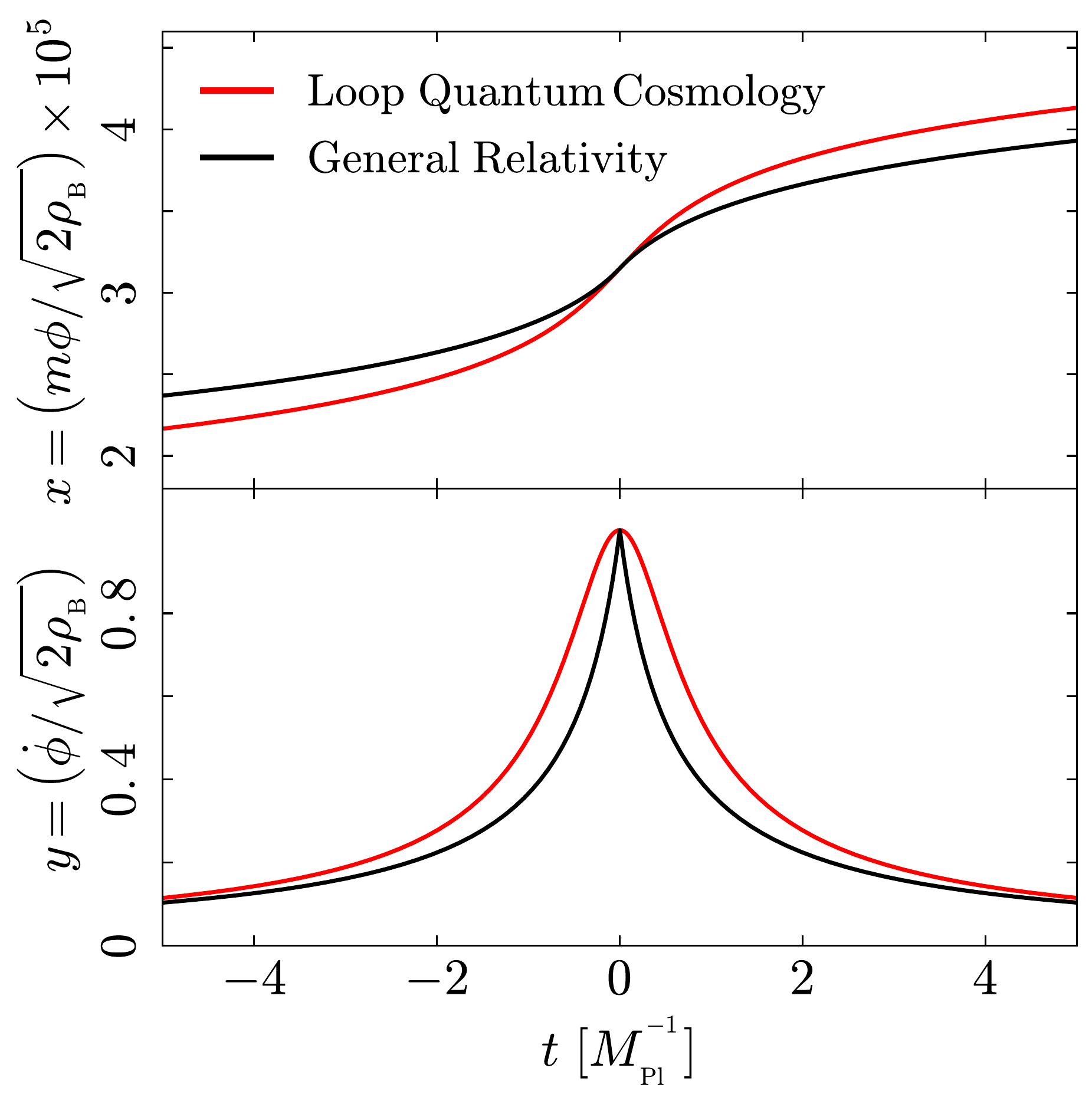}
\caption{The fraction of potential and kinetic energy in general relativity (black) and loop quantum cosmology (blue). }
\label{Hgrlqc}
\end{figure}

We shall now investigate to which extent the specific modified dynamics is responsible for the peakedness of the probability density function of the number of e-folds in loop quantum cosmology. The argument we have developed in the previous section did not refer to the modified LQC dynamics. It was essentially based on Gibbons and Turok's analysis combined with the presence of the bounce at Planckian energy density. It can therefore already be guessed that the peakedness does not depend strongly on the LQC modification to the Friedmann equation. To address this question in more details, we consider an artificial bouncing cosmological scenario where the Friedmann equation is left unchanged even at Planckian energy. In this ``GR-like" cosmological scenario, initial conditions for a given trajectory are set in the remote past of the contracting branch at the same energy density and with the same values of $\qsubrm{\phi}{0}$ and $\qsubrm{\dot{\phi}}{0}$ than for a trajectory which follows the LQC dynamics (as previously considered). The dynamics is divided into two parts: the contracting branch with a negative Hubble parameter and the expanding branch with a positive Hubble parameter. The evolution, starting in the contracting branch, is artificially stopped when the energy density reaches the LQC critical energy density $\qsubrm{\rho}{B}$. The values of $\qsubrm{\phi}{B}$ and $\qsubrm{\dot{\phi}}{B}$ are collected and used as initial conditions for the dynamics in the expanding branch where the initial Hubble parameter is now positive. At the junction between both phases, the Hubble parameter and therefore $\ddot{\phi}$ are discontinuous but $a$, $\phi$ and $\dot{\phi}$ are continuous, as illustrated in Fig. \ref{xlambda}, \ref{aandH} and \ref {Hgrlqc}. \\
The numerical result for the PDF of the number of e-folds in the GR-like scenario is plotted against the LQC prediction on the left panel of Fig. \ref{pdf-lqc}. The PDF has the same width and shape than in LQC. This confirms that the peakedness does not depend strongly on the specific LQC modified dynamics, and suggests that this feature would remain in case one incorporates additional quantum gravity corrections to the LQC effective equations. Nevertheless, the number of e-folds corresponding to the peak of the PDF is slightly different in the GR-like scenario than in LQC. This can be explained as follows.\\ 
First, it should be noticed that the difference between the GR-like scenario and LQC becomes significant when $\rho\approx\qsubrm{\rho}{Pl}$. 
Second, as shown in the previous sections, the fraction of trajectories that have a significant phase of deflation in the contracting branch is tiny.  This means that at high energy density, the dynamics of most trajectories is largely kinetic energy dominated. In simplistic terms, deflation can not bring the amplitude of the scalar field to large values because it stops nearly immediately. To inverstigate the difference between the GR-like scenario and LQC, the equation of motion of the scalar field at high energy, and for kinetic energy domination, need to be studied. It is natural to introduce $x\equiv\phi/\sqrt{2\qsubrm{\rho}{B}}$ and $y\equiv\dot{\phi}/\sqrt{2\qsubrm{\rho}{B}}$, the so-called potential and kinetic energy parameters. As the duration of inflation depends on the scalar field amplitude, it is sufficient to focus on the potential energy parameter. It is easy to show that in the interesting regime
\bse
\bea
\qsuprm{x}{GR}&\simeq&\qsubrm{x}{B}+(3/\Gamma)\ln a,\\
\qsuprm{x}{LQC}&\simeq&\qsubrm{x}{B}+(3/\Gamma)\ln{a} +(\ln 2/\Gamma).
\eea
\ese
In LQC the scalar field is \textit{boosted} by a short phase of super-inflation, $\dot{H}>0$, during which its amplitude accumulates a surplus of $(\ln 2/\Gamma)$ compared to the standard FLRW dynamics. This yields a difference of $\sqrt{8/3}\ln2\sqrt{\qsubrm{N}{GR}}$ between the number of e-folds in both scenarios. With the standard numerical values of $m$ and $\qsubrm{\rho}{B}$, one gets $\qsubrm{N}{LQC}-\qsubrm{N}{GR}\simeq13$,  in agreement with the numerical results. \\

\begin{figure}
\includegraphics[height=8.cm]{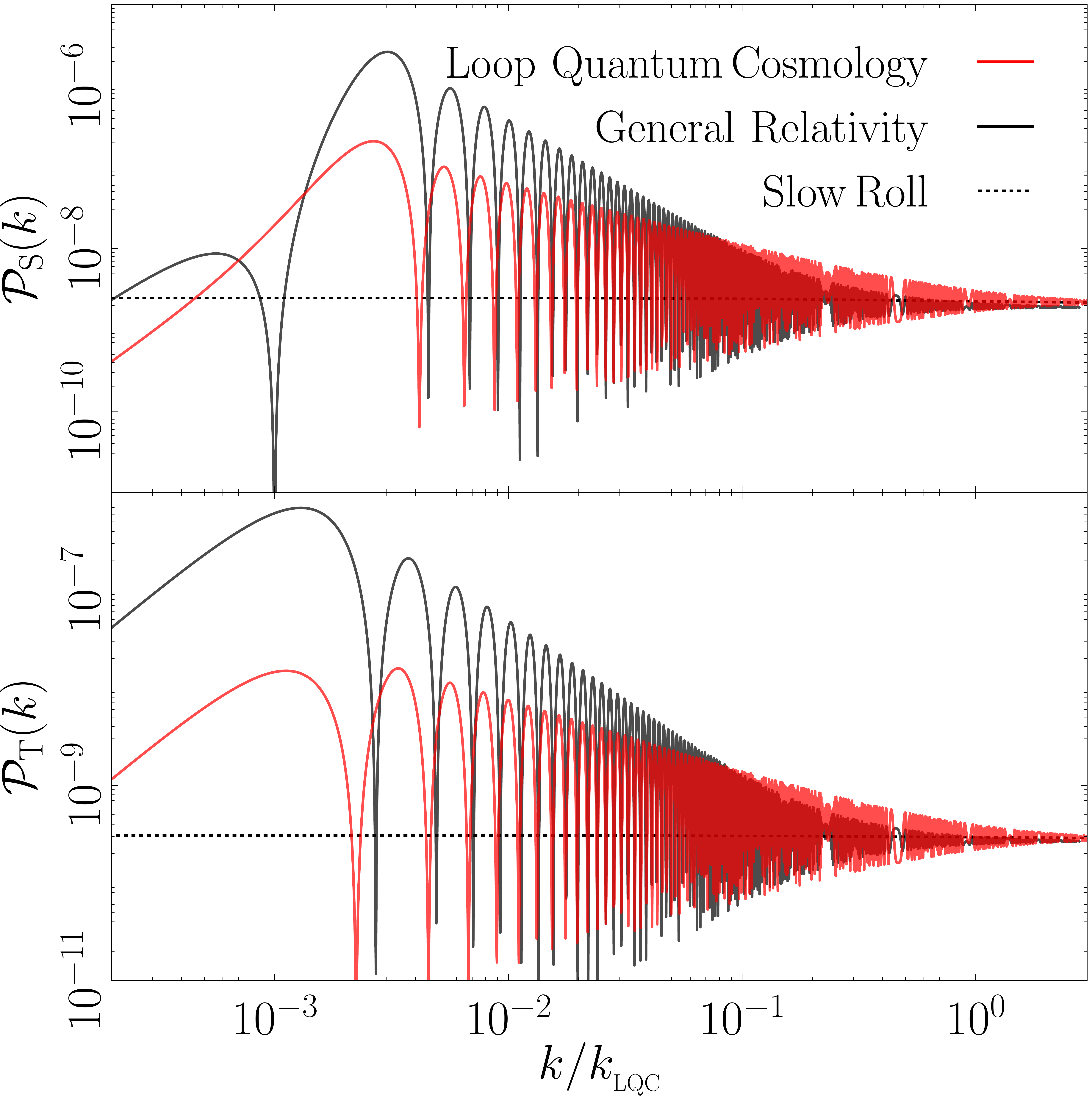}
\caption{Primordial power spectra of scalar (top) and tensor (bottom) perturbations for the GR-like dynamics in black (standard Friedmann equation, initial condition at the energy density corresponding to the energy density of the LQC bounce) and LQC dynamics in red (initial conditions at the bounce) plotted against the slow-roll expectation (dotted lines). On the the x-axis, $\qsubrm{k}{LQC}=\qsubrm{a}{B}\sqrt{\qsubrm{\rho}{B}}$.} 
\label{pspt}
\end{figure}
One can also study the differences between the primordial power spectra of cosmological perturbations in the GR-like scenario and in LQC. As a toy model to focus on the difference between both background dynamics, we set initial conditions for perturbations at an energy density corresponding to the bounce energy density, choosing the Bunch-Davies vacuum as the initial state. The resulting power spectra are shown on Fig. \ref{pspt} and compared with the usual slow-roll inflation expectation (dotted lines). Such spectra (and their variants including ore subtle LQC effects) are the main observables associated with loop quantum cosmology. \\
The length of inflation is crucial because it determines the location of the window of wavenumbers relevant for the cosmic microwave background anisotropy measurements, with respect to the characteristic LQC scale $\qsubrm{k}{LQC}\equiv\qsubrm{a}{B}\sqrt{\qsubrm{\rho}{B}}$.  On infrared scales, the mode functions remain in the Bunch-Davies state with $\mathcal{P}(k)\propto k^2$. We see that in both LQC and the GR-like scenario the power spectra agree with the slow-roll expectations in the ultraviolet regime. Oscillations are present in both scenarios in the range $10^{-3}<k/\qsubrm{k}{LQC}<1$. The amplitude is larger in the GR-like scenario than in LQC, however the period of the oscillations does not seem to be affected by the specific modified LQC dynamics. This shows that oscillations in themselves are a bounce feature but not a specific LQC feature. This motivates the search for complementary probes such as primordial non-gaussianity \cite{Agullo:2015aba}. For a more detailed comparison of the different kinds of power spectra expected in LQC under different assumptions for the mode propagation, see \cite{Bolliet:2015bka}.

The most reliable result of loop quantum cosmology is the modified Friedmann equation describing the background dynamics. It receives a quadratic correction in density which prevents the Universe from collapsing into a singularity. In this article, we have investigated the influence of this modified dynamics on the duration of inflation. The conclusion is that the modification of the Friedmann equation has a very small (but non vanishing) impact on the duration of inflation. The key role played by LQC in ``predicting" inflation -- or more precisely the duration of inflation -- is not due to the modified dynamics in itself. It is grounded in two different aspects. First, LQC sets the energy scale. This is the fundamental point. As far as the Universe is assumed to be filled by a massive scalar field, inflation happens naturally if the energy scale ``before" inflation is high enough, $\rho\gg m^2$. But whereas starting at the Planck energy density in GR is somehow arbitrary, in LQC the bounce energy density can be calculated (modulo some hypotheses) and derived from the full theory, providing a natural energy scale. This is the first important aspect. Second, LQC selects favored conditions at the bounce, see formula \eqref{eq:gt}, corresponding to a favored duration of inflation $N\simeq145$, for $\qsubrm{\rho}{B}=0.41\qsubrm{m}{Pl}^4$. This is an interesting prediction rooted in the existence of a pre-bounce phase where a natural variable to which a known PDF can be assigned was identified. This cannot be produced in standard cosmology and is specific to bouncing models.

\section{Acknowledgement}

The work of BB  was supported by a grant from ENS Lyon and a Fulbright Grant from the Franco-American commission. 
\bibliography{refs}

\begin{thebibliography}{7}
\expandafter\ifx\csname natexlab\endcsname\relax\def\natexlab#1{#1}\fi
\expandafter\ifx\csname bibnamefont\endcsname\relax
  \def\bibnamefont#1{#1}\fi
\expandafter\ifx\csname bibfnamefont\endcsname\relax
  \def\bibfnamefont#1{#1}\fi
\expandafter\ifx\csname citenamefont\endcsname\relax
  \def\citenamefont#1{#1}\fi
\expandafter\ifx\csname url\endcsname\relax
  \def\url#1{\texttt{#1}}\fi
\expandafter\ifx\csname urlprefix\endcsname\relax\def\urlprefix{URL }\fi
\providecommand{\bibinfo}[2]{#2}
\providecommand{\eprint}[2][]{\url{#2}}

\bibitem[{\citenamefont{{Ashtekar} and {Singh}}(2011)}]{2011CQGra..28u3001A}
\bibinfo{author}{\bibfnamefont{A.}~\bibnamefont{{Ashtekar}}} \bibnamefont{and}
  \bibinfo{author}{\bibfnamefont{P.}~\bibnamefont{{Singh}}},
  \bibinfo{journal}{Classical and Quantum Gravity}
  \textbf{\bibinfo{volume}{28}}, \bibinfo{eid}{213001} (\bibinfo{year}{2011}),
  \eprint{1108.0893}.

\bibitem[{\citenamefont{{Gibbons} and {Turok}}(2008)}]{2008PhRvD..77f3516G}
\bibinfo{author}{\bibfnamefont{G.~W.} \bibnamefont{{Gibbons}}}
  \bibnamefont{and} \bibinfo{author}{\bibfnamefont{N.}~\bibnamefont{{Turok}}},
  \bibinfo{journal}{\prd} \textbf{\bibinfo{volume}{77}}, \bibinfo{eid}{063516}
  (\bibinfo{year}{2008}), \eprint{hep-th/0609095}.

\bibitem[{\citenamefont{Ashtekar and Sloan}(2011)}]{AS2011}
\bibinfo{author}{\bibfnamefont{A.}~\bibnamefont{Ashtekar}} \bibnamefont{and}
  \bibinfo{author}{\bibfnamefont{D.}~\bibnamefont{Sloan}},
  \bibinfo{journal}{Gen. Rel. Grav.} \textbf{\bibinfo{volume}{43}},
  \bibinfo{pages}{3619} (\bibinfo{year}{2011}), \eprint{1103.2475}.

\bibitem[{\citenamefont{Linsefors and Barrau}(2013)}]{Linsefors:2013cd}
\bibinfo{author}{\bibfnamefont{L.}~\bibnamefont{Linsefors}} \bibnamefont{and}
  \bibinfo{author}{\bibfnamefont{A.}~\bibnamefont{Barrau}},
  \bibinfo{journal}{Phys. Rev.} \textbf{\bibinfo{volume}{D87}},
  \bibinfo{pages}{123509} (\bibinfo{year}{2013}), \eprint{1301.1264}.

\bibitem[{\citenamefont{Linsefors and Barrau}(2015)}]{Linsefors:2014tna}
\bibinfo{author}{\bibfnamefont{L.}~\bibnamefont{Linsefors}} \bibnamefont{and}
  \bibinfo{author}{\bibfnamefont{A.}~\bibnamefont{Barrau}},
  \bibinfo{journal}{Class. Quant. Grav.} \textbf{\bibinfo{volume}{32}},
  \bibinfo{pages}{035010} (\bibinfo{year}{2015}), \eprint{1405.1753}.

\bibitem[{\citenamefont{Agullo}(2015)}]{Agullo:2015aba}
\bibinfo{author}{\bibfnamefont{I.}~\bibnamefont{Agullo}},
  \bibinfo{journal}{Phys. Rev.} \textbf{\bibinfo{volume}{D92}},
  \bibinfo{pages}{064038} (\bibinfo{year}{2015}), \eprint{1507.04703}.

\bibitem[{\citenamefont{Bolliet et~al.}(2015)\citenamefont{Bolliet, Grain,
  Stahl, Linsefors, and Barrau}}]{Bolliet:2015bka}
\bibinfo{author}{\bibfnamefont{B.}~\bibnamefont{Bolliet}},
  \bibinfo{author}{\bibfnamefont{J.}~\bibnamefont{Grain}},
  \bibinfo{author}{\bibfnamefont{C.}~\bibnamefont{Stahl}},
  \bibinfo{author}{\bibfnamefont{L.}~\bibnamefont{Linsefors}},
  \bibnamefont{and} \bibinfo{author}{\bibfnamefont{A.}~\bibnamefont{Barrau}},
  \bibinfo{journal}{Phys.Rev.} \textbf{\bibinfo{volume}{D91}},
  \bibinfo{pages}{084035} (\bibinfo{year}{2015}), \eprint{1502.02431}.

\end{thebibliography}
 \end{document}